\documentclass[twocolumn]{aastex631}

\usepackage{amsmath}
\newcommand{\pfsspy}{\texttt{pfsspy} }

\begin{document}

\title{Test Problems for Potential Field Source Surface Extrapolations of Solar and Stellar Magnetic Fields}
\author[0000-0002-1365-1908]{David Stansby}
\affiliation{Mullard Space Science Laboratory, University College London, Holmbury St. Mary, Surrey RH5 6NT, UK}
\author[0000-0002-0497-1096]{Daniel Verscharen}
\affiliation{Mullard Space Science Laboratory, University College London, Holmbury St. Mary, Surrey RH5 6NT, UK}

\begin{abstract}
The potential field source surface (PFSS) equations are commonly used to model the coronal magnetic field of the Sun and other stars. As with any computational model, solving equations using a numerical scheme introduces errors due to discretisation. We present a set of tests for quantifying these errors by taking advantage of analytic solutions to the PFSS equations when the input field is proportional to a single spherical harmonic. From the spherical harmonic solutions we derive analytic equations for magnetic field lines traced through the three dimensional magnetic field solution. We propose these as a set of standard analytic solutions that all PFSS solvers should be tested against to quantify their inherent errors. We apply these tests to the \pfsspy software package, showing that it reproduces spherical harmonic solutions well with a slight overestimation of the unsigned open magnetic flux. It is also successful at reproducing analytic field line equations, with errors in field line footpoints typically much less than one degree.
\end{abstract}


\section{Introduction}
The potential field source surface (PFSS) set of equations \citep{Altschuler1969, Schatten1969} are commonly used to model the coronal magnetic field of the Sun \citep[e.g.][]{Badman2020, Stansby2021, Fargette2021} and other stars \citep[e.g.][]{Jardine2017, Saikia2020, Kochukhov2020}. The key assumption of PFSS models is the absence of electric current within the domain of interest. Under this assumption the electromagnetic equations for the magnetic field ($\mathbf{B}$) reduce to
\begin{align}
	\nabla \cdot \mathbf{B} &= 0\\
	\nabla \times \mathbf{B} &= \mathbf{0}
\end{align}
These equations are equivalent to the Laplace equation,
\begin{equation}
	\nabla^{2} \Phi = 0,
\end{equation}
where the scalar potential is defined as $\mathbf{B} = \nabla \Phi$. The solution is derived between the stellar surface (at a radius r$_{\odot}$) and a specified source surface (at a radius $r_{ss} > r_{\odot}$), and three boundary conditions are required for a unique solution. One of these is given by a single component of the magnetic field at the inner boundary ($r = r_{\odot}$), and the other two are set by the assumption that the field is purely radial (ie. the two transverse components are zero) on the outer boundary ($r = r_{ss}$). The second assumption is motivated by accelerating stellar wind plasma forcing the field to be near-radial at the source surface. 

Different numerical methods are available for solving the above equations, including spherical harmonic expansions \citep[e.g.][]{Altschuler1969, Hakamada1995, Toth2011} and finite difference methods \citep[e.g.][]{Jiang2012, Toth2011, Caplan2021}. To understand the topology of the magnetic field \citep[e.g.][]{vanDriel-Gesztelyi2012, Boe2020, Baker2021} and how stellar wind flows through a corona \citep[e.g.][]{Neugebauer1998, Stansby2020f}, magnetic field line connectivities between the outer and inner boundaries are also needed, which requires tracing magnetic field lines through the three-dimensional magnetic field solution using a field line tracer.

In order to check whether numerical methods work as expected, and to quantify any errors inherent in the numerical scheme employed, it is helpful to compare their output with exact analytical solutions. While in some areas such as hydrodynamic modelling test problems are common \citep{Sod1978}, to our knowledge no such test problems have been published for PFSS solvers and field line tracers.

In this paper we provide a set of analytical solutions for the PFSS equations (Section \ref{sec:pfss}) and use these to derive solutions for the unsigned open flux at the source surface (Section \ref{sec:open flux}) and field lines traced through these solutions (Section \ref{sec:streamlines}).  We then compare these to the numerical solutions computed by the \pfsspy solver \citep{Stansby2020e} (Section \ref{sec:comparison}), to demonstrate its usefulness, accuracy, and limitations.

\section{Analytical solutions}
In this section we recount analytical solutions to the PFSS equations using spherical harmonics (Section \ref{sec:pfss}). Using these solutions we then derive equations for the total open magnetic flux (Section \ref{sec:open flux}) and magnetic field lines traced through the spherical harmonic solutions (Section \ref{sec:streamlines}).

\subsection{PFSS solutions}
\label{sec:pfss}
Following \cite{Wang1992, Mackay2012}, when the radial component of the magnetic field on the stellar surface ($r_{\odot}$) is specified the general solution to the PFSS equations in a spherical coordinate system is given by the spherical harmonic decomposition
\begin{align}
	B_{r} &= \sum_{\ell=1}^{\infty} \sum_{m=-\ell}^{m=\ell} a_{\ell m} c_{\ell} \left (r \right ) Y_{\ell m} \left (\theta, \phi \right ) \label{eq:br} \\
	B_{\theta} &= \sum_{l=1}^{\infty} \sum_{m=-\ell}^{m=\ell} a_{\ell m} d_{\ell} \left (r \right ) \frac{\partial Y_{\ell m} \left (\theta, \phi \right )}{\partial \theta} \\
	B_{\phi} &= \sum_{\ell=1}^{\infty} \sum_{m=-\ell}^{m=\ell}  a_{\ell m} d_{\ell} \left (r \right ) \frac{1}{\sin \theta} \frac{ \partial Y_{\ell m} \left (\theta, \phi \right )}{\partial \phi}
	\label{eq:bphi}
\end{align}
where
\begin{equation}
	Y_{\ell m} \left ( \theta, \phi \right ) = A_{\ell m} P_{\ell m} \left ( \cos \theta \right ) \begin{Bmatrix} \cos m\phi\\  \sin m\phi \end{Bmatrix}~~~~~~~~~~  \begin{matrix} m > 0 \\  m \leq 0 \end{matrix}
\end{equation}
are real spherical harmonics, $A_{\ell m}$ are normalisation coefficients and $P_{\ell m}$ are the associated Legendre polynomials.

$a_{\ell m}$ are constant coefficients derived from the input magnetic field via.
\begin{equation}
	a_{\ell m} = \int_{0}^{2\pi} d\phi \int_{0}^{\pi} \sin \theta d\theta ~B_{r \odot} \left ( \theta, \phi \right ) Y_{\ell m} \left ( \theta, \phi \right )
\end{equation}
and the functions $c_{\ell}$ and $d_{\ell}$ are
\begin{align}
	c_{\ell} \left ( z \right ) &= \left ( \frac{r}{r_{\odot}} \right )^{-\ell - 2} \left [ \frac{\ell + 1 + \ell \left ( r / r_{ss} \right )^{2\ell + 1}}{\ell + 1 + \ell (r_{\odot} / r_{ss})^{2\ell + 1}} \right ] \\
	d_{\ell} \left ( z \right ) &= \left ( \frac{r}{r_{\odot}} \right )^{-\ell - 2}  \left [ \frac{1 -  \left ( r / r_{ss} \right )^{2\ell + 1}}{\ell + 1 + \ell (r_{\odot} / r_{ss})^{2\ell + 1}} \right ]
	\label{eq:d}
\end{align}
The polar coordinate range is $[0, \pi]$, with $\theta = 0$ the north pole.

If the input magnetic field is directly proportional to a single spherical harmonic, $B_{r} = B_{0} Y_{\ell'm'}$ where $B_{0}$ is a constant, and the $a_{\ell m}$ coefficients simplify greatly to
\begin{equation}
	a_{\ell m} = B_{0} \delta_{\ell \ell'} \delta_{mm'} 
\end{equation}
and only a single term in each sum is non-zero. For the remainder of this paper we set the input magnetic field proportional to a single spherical harmonic, so for brevity drop the apostrophes that denote a specific choice of $\ell,m$. Under these assumptions the solutions are
\begin{align}
	\label{eq:harm_solution_start}
	B_{r} &= B_{0} c_{\ell} \left (r \right ) Y_{\ell m} \left (\theta, \phi \right ) \\
	B_{\theta} &= B_{0} d_{\ell} \left (r \right ) \frac{\partial Y_{\ell m} \left (\theta, \phi \right )}{\partial \theta} \\
	B_{\phi} &= B_{0} d_{\ell} \left (r \right ) \frac{1}{\sin \theta} \frac{ \partial Y_{\ell m} \left (\theta, \phi \right )}{\partial \phi}
	\label{eq:harm_solution_end}
\end{align}
In the next sections we provide equations for calculating the total unsigned open flux (Section \ref{sec:open flux}) and magnetic field line equations (Section \ref{sec:streamlines}) from these solutions.
\subsection{Open flux}
\label{sec:open flux}
The total unsigned open flux is defined by integrating the radial component of the magnetic field on the source surface
\begin{equation}
	\Phi_{open} =  r^{2}_{ss} \int_{0}^{2\pi} \int_{0}^{\pi}  \left | B_{r} \left ( \theta, \phi, r_{ss} \right ) \right | \sin \theta d\theta d\phi
\end{equation}
For a single harmonic this simplifies to
\begin{align}
	\Phi_{open} = r^{2}_{ss} B_{0} c_{\ell} \left ( r_{ss} \right ) \int_{0}^{2\pi} \int_{0}^{\pi} \left | Y_{\ell m} \left (\theta, \phi \right ) \right | \sin \theta d\theta  d\phi
	\label{eq:open flux}
\end{align}
We numerically evaluated the double integral to get a number for the analytic open flux when comparing to values computed from the PFSS solver.

\subsection{Magnetic field lines}
\label{sec:streamlines}
In spherical coordinates the magnetic field tracing equations are
\begin{align}
	\frac{dr}{ds} &= \hat{B}_{r}\\
	\frac{d\theta}{ds} &= \frac{\hat{B}_{\theta}}{r}\\
	\frac{d\phi}{ds} &= \frac{\hat{B}_{\phi}}{r\sin\theta}
\end{align}
where the paramater $s$ is the physical distance along the field line, and $\hat{B}_{i} := B_{i} / \left | \mathbf{B} \right |$ are components of a unit vector pointing in the direction of the magnetic field. These form a set of three coupled equations that can be integrated from an initial seed point to evaluate coordinates along a magnetic field line.

In the case where the input field is proportional to a single spherical harmonic, equations \ref{eq:harm_solution_start} -- \ref{eq:harm_solution_end} can be substituted in and $ds$ eliminated to give
\begin{align}
	\frac{d\theta}{dr} &= \left [ \frac{1}{r} \frac{d_{\ell} \left ( r \right )}{c_{\ell} \left ( r \right )} \right ] \left [\frac{1}{Y_{\ell m}} \frac{\partial Y_{\ell m} \left (\theta, \phi \right )}{\partial \theta}  \right ] \label{eq:theta fline 0} \\
	\frac{d\phi}{dr} &= \frac{1}{r \sin^{2} \theta} \frac{d_{\ell} \left ( r \right )}{c_{\ell} \left ( r \right )}  \left [\frac{1}{Y_{\ell m}} \frac{\partial Y_{\ell m} \left (\theta, \phi \right )}{\partial \phi}  \right ] \\
	\frac{d \theta}{d \phi} &= \sin^{2} \theta \left [  \frac{\partial Y_{\ell m} \left (\theta, \phi \right )}{\partial \theta}  \right ] \left [  \frac{\partial Y_{\ell m} \left (\theta, \phi \right )}{\partial \phi}  \right ]^{-1} \label{eq:phi fline 0}
\end{align}
Because the spherical harmonics are separable in $\theta, \phi$, equation \ref{eq:theta fline 0} is a function of only $r$ and $\theta$ and equation \ref{eq:phi fline 0} is only a function of $\phi$ and $\theta$. This decouples the three field line tracing equations, allowing two of them to be integrated independently to give the longitude and latitude as functions of radius. In the following subsections we integrate these two equations and give analytic solutions for field lines in a spherical harmonic solution.

\subsubsection{The $\theta$ field line equation}
To find $\theta \left ( r \right )$ we start by separating variables and integrating Equation \ref{eq:theta fline 0}
\begin{equation}
	\int_{\theta_{ss}}^{\theta} P_{\ell m} \left ( \bar{\theta} \right ) \left [ \frac{\partial P_{\ell m} \left (\bar{\theta} \right )}{\partial \bar{\theta}} \right ]^{-1} d\bar{\theta} = \int_{r_{ss}}^{r} \frac{1}{\bar{r}} \frac{d_{\ell} \left ( \bar{r} \right )}{c_{\ell} \left ( \bar{r} \right )} d\bar{r}  \label{eq:theta fline}
\end{equation}
Barred symbols denote dummy integration variables. To trace field lines down from the source surface to the solar surface, the radial integration limits are set to the source surface radius, $r_{ss}$ and the radial coordinate along the field line, $r$. The $\theta$ integration limits set to a fixed initial latitude $\theta_{ss}$, and the latitude along the field line, $\theta$. Defining $\rho = r / r_{ss}$, the radial integration is given in equation 23 of \cite{Gregory2011}  as
\begin{equation}
	\int_{r_{ss}}^{r} \frac{1}{\bar{r}} \frac{d_{\ell} \left ( \bar{r} \right )}{c_{\ell} \left ( \bar{r} \right )} d\bar{r} = \frac{1}{\ell} \frac{1}{\ell+1} \ln \left [ \frac{\rho^{\ell}(2\ell + 1)}{l\rho^{2\ell+1} + (\ell+1)} \right ]
	\label{eq:r integral}
\end{equation}

The $\theta$ integration on the left hand side of equation \ref{eq:theta fline} is more complicated -- \cite{Gregory2011} give a general solution for the left hand side of \ref{eq:theta fline} $\forall \ell$ and $m=0$, but for $\ell>1$ the solution is not available in closed form. Instead we rearrange the integral on the left hand side of equation \ref{eq:theta fline} to define the function
\begin{equation}
	 f_{\ell m} \left (\theta \right ) := \exp \left [2l \int_{\theta_{ss}}^{\theta} P_{\ell m} \left (\bar{\theta} \right ) \left [ \frac{\partial P_{\ell m} \left ( \bar{\theta} \right )}{\partial \bar{\theta}} \right ]^{-1} d\bar{\theta}  \right ]
	 \label{eq:flm}
\end{equation}
With this definition, equations \ref{eq:r integral} and \ref{eq:flm} are substituted into equation \ref{eq:theta fline} to give the equation for $\theta \left (r \right )$ along a field line
\begin{equation}
	 f_{\ell m} \left (\theta \right ) = f_{\ell m} \left (\theta_{ss} \right )  \left [\frac{\rho^{\ell}(2\ell + 1)}{\ell\rho^{2\ell+1} + (\ell+1)} \right ]^{\frac{2}{\ell+1}}
	 \label{eq:general theta fline}
\end{equation}
The functions $f_{\ell m} \left (\theta \right )$ for low order spherical harmonics are tabulated in Table \ref{tab:harms}. Solving this equation analytically requires that the inverse of $f_{\ell m}$ exist in a closed form.

\begin{deluxetable}{cccc}
\tablenum{1}
\tablecaption{Low order associated Legendre polynomials, and related functions. See Equation \ref{eq:flm} for the definition of $f_{\ell m}$ \label{tab:harms} and Equation \ref{eq:glm} for the definition of $g_{\ell m}$. The functions listed are the same for $m \rightarrow -m$.}
\tablehead{\colhead{l, m} & \colhead{$P_{\ell m}$} & \colhead{$f_{\ell m} \left ( \theta \right )$} & \colhead{$g_{\ell m} \left ( \theta \right )$}}
\startdata
1, 0		& $\cos \theta$				& $\sin^{2} \theta$	 \\
1, 1		& $\sin \theta$				& $\cos^{2} \theta $ 	& $\frac{\sin \theta}{ \cos \theta}$ \\ \hline
2, 0		& $3 \cos^{2} \theta - 1$		& \\
2, 1		& $\cos \theta \sin \theta$		& $\cos 2\theta$	& $ \frac{\sin \theta}{ \sqrt{cos 2\theta}}$  	\\
2, 2		& $\sin^{2} \theta$			& $\cos^{2} \theta$ 			& $\left (\frac{\sin \theta}{\cos \theta}\right )^{2}$		\\ \hline
3, 0		& $5\cos^{3} \theta - 3\cos\theta$	& 	\\
3, 1		& $(5\cos^{2}\theta - 1)\sin\theta$	&  	\\
3, 2		& $\cos\theta \sin^{2} \theta$	& $3\cos(2\theta) + 1$	& $\frac{\sin^{2}\theta}{2 - 3\sin^{2} \theta}$	\\
3, 3		& $\sin^{3} \theta$			& $\cos^{2}\theta$				& $\frac{\sin \theta}{\cos\theta}$	\\
\enddata
\end{deluxetable}

\subsubsection{The $\phi$ field line equation}
Equation \ref{eq:phi fline 0} is the field line equation relating $\theta$ and $\phi$. For $m=0$, $B_{\phi} = 0$ and this has the trivial solution $\phi = const$, so we only consider the $m \neq 0$ case.

Separating variables and integrating gives
\begin{equation}
	\int_{\theta_{ss}}^{\theta} \frac{P_{\ell m}}{\sin \bar{\theta} ^{2}} \left [ \frac{\partial P_{\ell m}}{\partial \bar{\theta}} \right ]^{-1} d\bar{\theta} = \int_{\phi_{ss}}^{\phi} Y_{\ell m}\left [ \frac{\partial Y_{\ell m} }{\partial \bar{\phi}} \right ]^{-1} d\bar{\phi} \label{eq:phi fline}
\end{equation}
The $\phi$ integration on the right hand side evaluates as
\begin{equation}
	\int_{\phi_{ss}}^{\phi} Y_{\ell m}\left [ \frac{\partial Y_{\ell m} }{\partial \bar{\phi}} \right ]^{-1} d\bar{\phi} =  \frac{1}{m^{2}}  \begin{Bmatrix}\ln \frac{\sin m\phi_{ss}}{\sin m\phi} \\ \ln \frac{\cos m\phi_{ss}}{\cos m\phi} \end{Bmatrix}~~~~~~~~~~  \begin{matrix} m > 0 \\  m < 0 \end{matrix} \label{eq:phi integration}
\end{equation}
For low order spherical harmonics the integral on the left hand side of \ref{eq:phi fline} is rearranged to define the function
\begin{equation}
	g_{\ell m} \left ( \theta \right ) := \exp \left [ m^{2} \int_{\theta_{ss}}^{\theta} \frac{P_{\ell m}}{\sin \bar{\theta}^{2}} \left [ \frac{\partial P_{\ell m}}{\partial \bar{\theta}} \right ]^{-1} d\bar{\theta} \right ]
	\label{eq:glm}
\end{equation}
The functions $g_{\ell m}$ for low order spherical harmonics are given in Table \ref{tab:harms}. Substituting equations \ref{eq:phi integration} and \ref{eq:glm} into \ref{eq:phi fline} gives the the equation for $\phi$ as a function of $r$ along a field line
\begin{equation}
	\begin{Bmatrix} \frac{\sin m\phi}{\sin m\phi_{ss}} \\ \frac{\cos m\phi}{\cos m\phi_{ss}} \end{Bmatrix} = \frac{g_{\ell m} \left ( \theta \right )}{ g_{\ell m} \left ( \theta_{ss} \right )}  ~~~~~~~~~~  \begin{matrix} m > 0 \\  m < 0 \end{matrix}
	\label{eq:general phi fline}
\end{equation}
Once $\theta (r)$ is known from solving equation \ref{eq:general theta fline}, it is used in Equation \ref{eq:general phi fline} to solve for $\phi (\theta (r))$. Unlike $f_{\ell m}$, the inverse of $g_{\ell m}$ does not need to exist in closed form to derive $\phi$ along a field line.

\section{The \pfsspy solver}
We briefly recount how \pfsspy calculates the PFSS solution, in order to understand the user configurable options that affect the accuracy of the solver.  Details on the numerical scheme are available in the numerical methods document archived alongside the software \citep{Stansby2022}. 

\pfsspy uses a finite difference method \citep[][Appendix B]{Ballegooijen2000} to calculate the magnetic vector potential on a grid regularly spaced in $z = \log \left ( r / r_{\odot} \right )$, $\lambda = \cos \theta$, $\phi$. The number of grid points in the angular dimensions, $n_{\phi}$ and $n_{\theta}$, are fixed by the resolution of the input $B_{r\odot}$ grid which must span the full sphere. Since the smallest magnetograms widely used from the Sun have a grid size of of 360 $\times$ 180 in (longitude, latitude), for simplicity we keep this as the fixed angular grid size throughout the tests. The number of radial grid points, $n_{r}$ is user configurable.

To trace magnetic field lines \pfsspy offers two different field line tracers. Here we perform comparisons with the FORTRAN implementation\footnote{\url{https://github.com/dstansby/streamtracer}}, which is the fastest of the two.

The field line equations in the coordinates that \pfsspy uses are
\begin{align}
	\frac{d\rho}{ds} &= \hat{B}_{\rho} \\
	\frac{d\lambda}{ds} &= \sqrt{1 - \lambda^{2}} \hat{B}_{\lambda} \\
	\frac{d\phi}{ds} &= \frac{1}{ \sqrt{1 - \lambda^{2}}} \hat{B}_{\phi}
\end{align}
In the field line tracer three equations are integrated numerically using a 4th order Runge-Kutta method with a fixed discrete step size, $\Delta s$, which is specified by the user as a fraction of the grid cell size in the radial direction,  $\Delta r = \log (r_{ss} / r_{\odot}) / n_{r}$. Because the step size $\Delta s$ is kept constant relative to the log-scaled grid, larger physical steps are taken further away from the solar surface at larger values of $r$.

\begin{deluxetable}{cccc}
\tablenum{2}
\tablecaption{Configurable parameters in the \pfsspy solver and their values for each figure.\label{tab:parameters}}
\tablehead{\colhead{} & \colhead{$n_{\phi}, n_{\theta}$} & \colhead{$n_{r}$} & \colhead{$\Delta s$}}
\startdata
Figure \ref{fig:sphharm}		& (360, 180)				& 40	& 1	 	\\
Figure \ref{fig:open_flux}		& (360, 180)				& Varied	& 1	\\
Figure \ref{fig:field line map}	& (360, 180)				& 40	& 1	 	\\
Figure \ref{fig:step size}		& (360, 180)				& 40	& Varied	\\
\enddata
\end{deluxetable}
\section{Analytic solutions as test problems}
\label{sec:comparison}
In this section we compare the magnetic field on the source surface (Section \ref{sec:Bss}), the open flux (Section \ref{sec:flux comparison}), and the field line connectivity (Section \ref{sec:field line comparison}) between analytic and numerical \pfsspy solutions. For clarity analysis is limited to harmonics with $\ell \leq 5$. For the Sun these are the dominant harmonics at all times in the solar cycle \citep{DeRosa2012}.

A range of software is used for creating comparisons, including pfsspy \citep{Stansby2020e}, numpy \citep{Harris2020}, scipy \citep{Virtanen2020a}, pandas \citep{Reback2021}, sympy \citep{Meurer2017} and astropy \citep{TheAstropyCollaboration2018}. Code for producing the comparisons is available in version 1.1.0 of the \pfsspy repository\footnote{\url{https://github.com/dstansby/pfsspy}} and is archived at \cite{Stansby2022}.

In the numerical solutions we set default the grid sizes to $n_{\phi} = 360$, $n_{\theta} = 180$, $n_{r} = 40$, and the default field line tracing step size to $\Delta s = 1$. The following subsections show a series of different tests, where one of these parameters may be varied. In all tests the source surface height is set to $r_{ss} = 2r_{\odot}$. Each test has one or more corresponding figures. For a summary of which parameters are fixed or varied for each figure, see Table \ref{tab:parameters}.

\subsection{$\mathbf{B}$ on the source surface}
\label{sec:Bss}
\begin{figure*}
\plotone{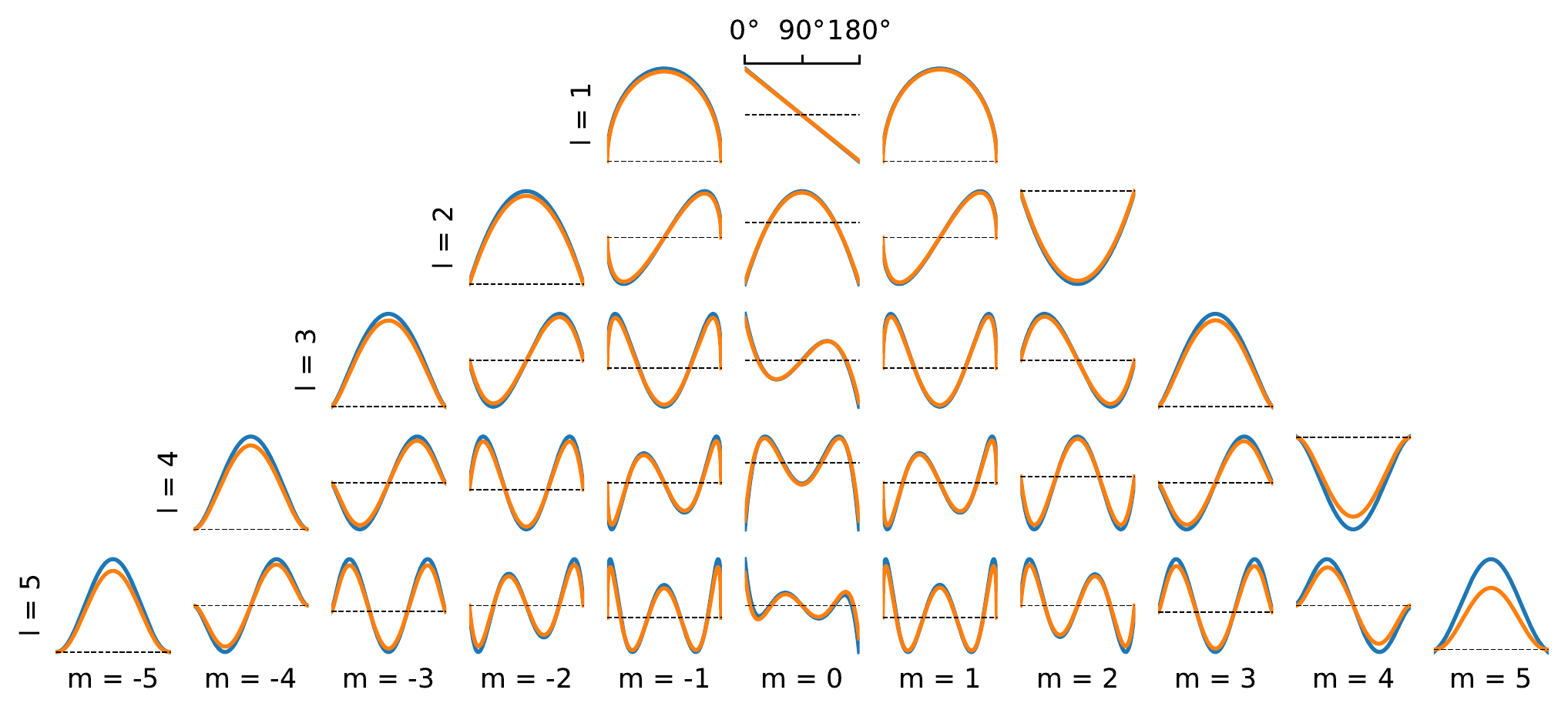}
\caption{Comparison between analytic (orange) and \pfsspy (blue) spherical harmonic solutions to the PFSS equations as functions of latitude at $r=r_{ss}, \phi=15^{\circ}$. Each panel represents a different spherical harmonic number, and plotted is the radial component of magnetic field. Dashed horizontal lines show $B_{r} = 0$. \label{fig:sphharm}}
\end{figure*}
The two transverse magnetic field components, $B_{\phi, \theta}$ must always be zero on the source surface within a PFSS solution. This is always the case in in \pfsspy which forces $B_{\phi, \theta} = 0$ at $r = r_{ss}$.

To compare $B_{r}$ at $r = r_{ss}$, equation \ref{eq:harm_solution_end} is evaluated at the same points as the numerical \pfsspy solution. 1D cuts of both the analytic and \pfsspy solutions at a constant longitude of $\phi=15^{\circ}$ are compared in Figure \ref{fig:sphharm}. \pfsspy reproduces the analytic solutions well, with the magnitude of solutions slightly larger than the analytic solutions. This suggests that \pfsspy systematically over-estimates the total unsigned open flux, which we investigate quantitively in the next section.

\subsection{Open flux}
\label{sec:flux comparison}
 
\begin{figure*}
\plotone{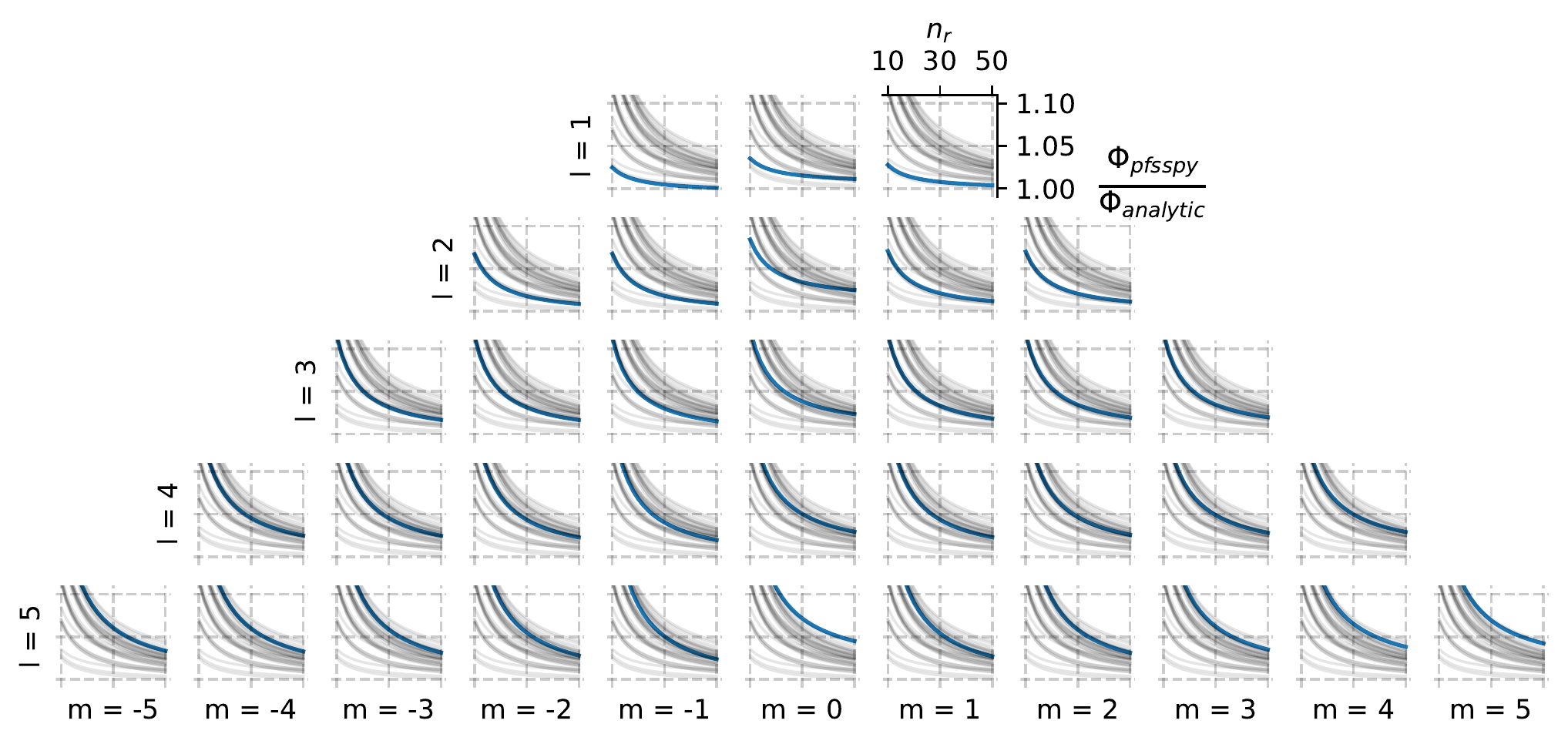}
\caption{Ratio of unsigned open flux in numerical \pfsspy solutions to analytic solutions as a function of number of radial cells within the \pfsspy solution for various spherical harmonics.
\label{fig:open_flux}}
\end{figure*}
Equation \ref{eq:open flux} gives an analytic integral for the unsigned open magnetic flux, which we evaluate using the \texttt{nquad} integration function in \texttt{scipy}. To check the accuracy of the integration we successfully verify the result against the analytical solution to the integral for $\ell,m = 1,0$:
\begin{align}
	\Phi_{open} / B_{0} &= c_{1} \left ( r_{ss} \right ) \int_{0}^{2\pi} d\phi \int_{0}^{\pi} \left | \cos\theta \right |\sin\theta d\theta \\
	&= \frac{3}{2z_{ss}^{3} + 1} \cdot 2\pi \cdot 2\int_{0}^{\pi/2} \cos\theta \sin\theta d\theta \\
	&= \frac{6\pi}{2z_{ss}^{3} + 1}
\end{align}

To evaluate the total unsigned open flux within a \pfsspy result the magnetic field values on the source surface is multiplied by the area of their associated cell and summed over the full sphere.

Figure \ref{fig:open_flux} shows the ratio of open flux in \pfsspy results to the analytic value as a function of the number of radial grid points for spherical harmonics up to $\ell=5$. As hinted from Figure \ref{fig:sphharm}, the unsigned open fluxes within \pfsspy are systematically larger than the analytic solution. This difference decreases with an increasing number of radial grid cells, approaching asymptotic values of $\leq 5\%$ at around $n_{r} = 40$. This motivates our choice of $n_{r} = 40$ for the default number of radial grid cells -- with more radial cells there is not a significant reduction in the open flux error.

At larger values of $\ell$ the error in unsigned open flux increases from $\sim$1\% at $\ell =1$ to $\sim$5\% at $\ell = 5$. The contribution of each spherical harmonic to the open flux on the source surface reduces as $(r_{ss} / r_{\odot})^{-\ell-2}$ (Equation \ref{eq:d}). As long as the spherical harmonics of the input $B_{r}$ map do not scale exponentially with $\ell$, the increasing open flux error with $\ell$ within \pfsspy will be suppressed by the $(r_{ss} / r_{\odot})^{-\ell-2}$ factor, preventing the total open flux error growing when summing over multiple harmonics.

\subsection{Field line connectivity}
\label{sec:field line comparison}
\begin{figure*}
\plotone{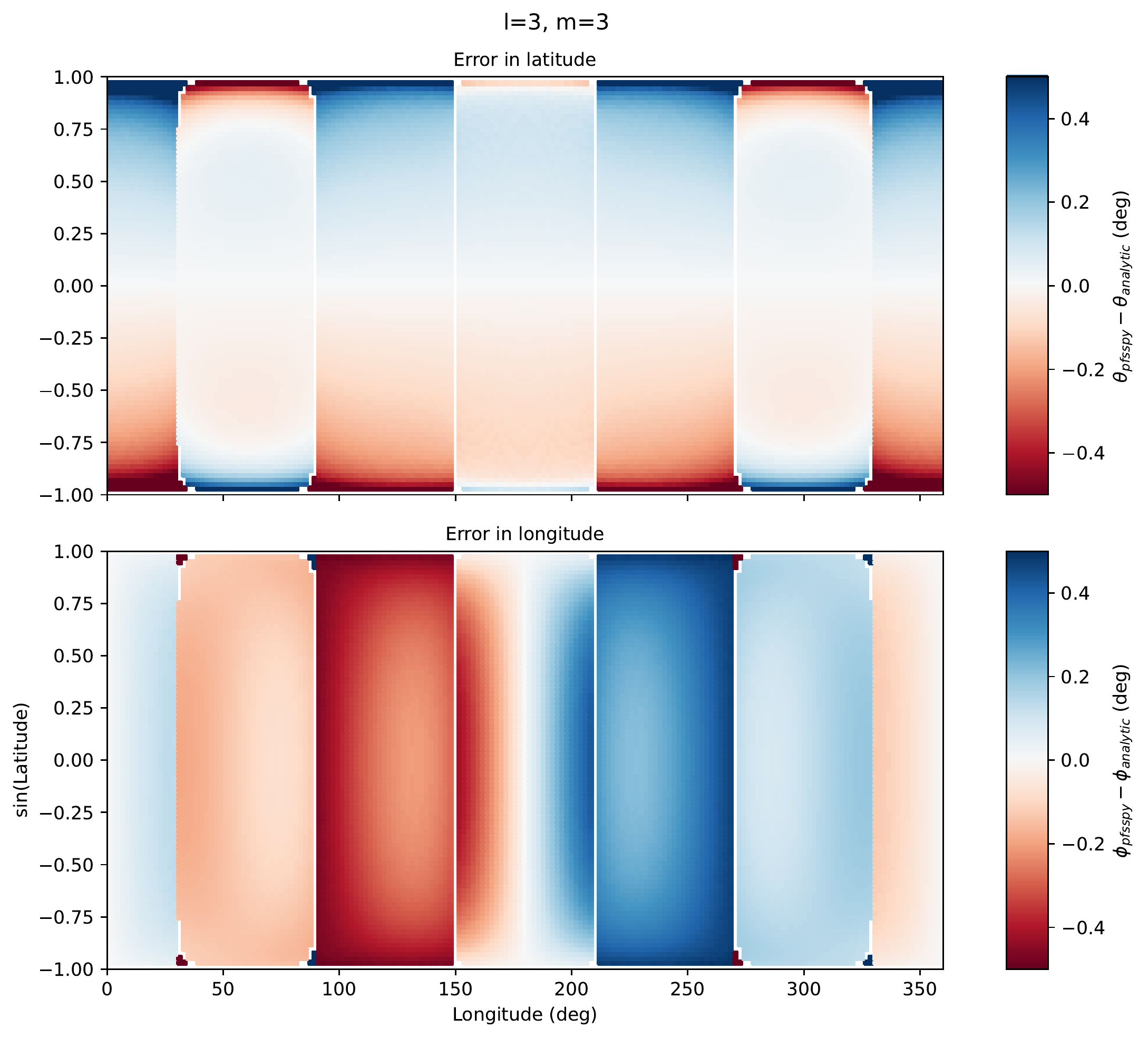}
\caption{The difference between numerically traced and analytic solar surface footpoint latitude (top panel) and longitude (bottom panel), for $l=3, m=3$ and a tracer step size of $\Delta s =1$. Field lines are traced downwards from an evenly spaced grid on the source surface.
\label{fig:field line map}}
\end{figure*}

From a regularly spaced grid of field line seed points on the source surface analytic solar surface field line footpoints are calculated using Equations \ref{eq:general theta fline} and \ref{eq:general phi fline} and the numerical field line footpoints computed using \pfsspy.  Figure \ref{fig:field line map} shows the difference between the analytic and traced solar surface footpoints in latitude (top panel) and longitude (lower panel) for $\ell,m = 3,3$ a step size of $\Delta s = 1$, and a radial grid size of $n_{r} = 40$. Overall the errors are small in both latitude and longitude at $\leq 0.5^{\circ}$.

There are two situations where \pfsspy fails to correctly trace the field lines. The first is near polarity inversion lines, which occur at $30^{\circ}, 90^{\circ}, 150^{\circ},...$ in Figure \ref{fig:field line map}. Traced field lines started near polarity inversion lines on the source surface turn around and return immediately to the source surface. This is due to finite resolution effects within the model. \pfsspy automatically tags these field lines as incorrectly traced, and they show up as thin white strips in Figure \ref{fig:field line map}. This occurs at all spherical harmonic numbers, but only for a limited number of field lines near polarity inversion lines.

The second situation occurs near polarity inversion lines at high latitudes, when the inversion line in the model is slightly displaced from the expected inversion line. This causes field lines to deviate strongly in longitude from their analytic solution, as seen in the bottom panel of Figure \ref{fig:field line map} where a few points have large errors outside the range of the errorbars. This is due to a combination of finite resolution effects and no attempt by the tracer to handle the spherical coordinate singularity at the poles. Up to $\ell=3$ it only occurs for $\ell=3$ and $m=-3, 3$.

\begin{figure*}
\plotone{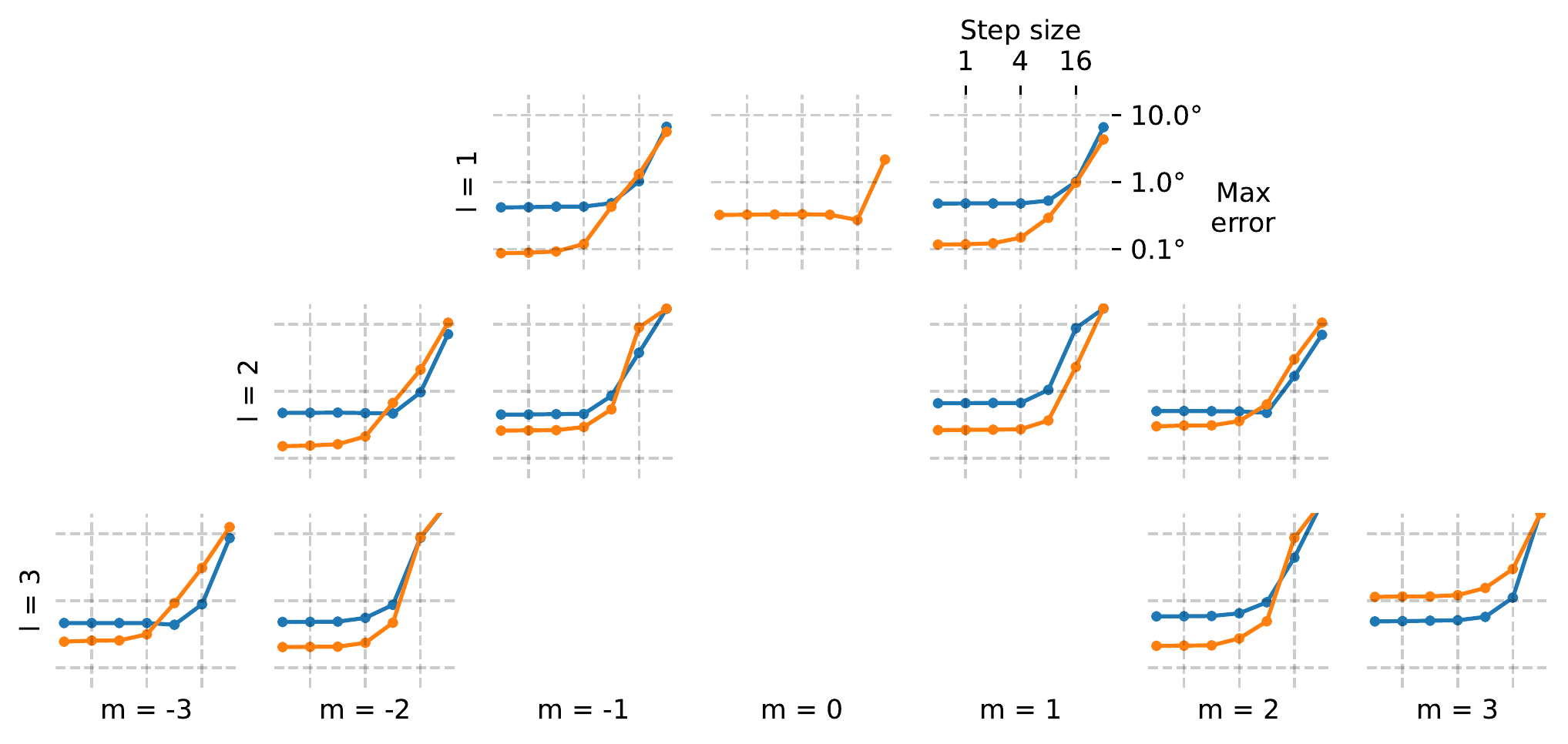}
\caption{Maximum error in field line footpoints as a function of tracer step size for various spherical harmonics. Error in longitude is shown in blue and error in latitude in orange. Errors with absolute values $> 10^{\circ}$ are excluded in these plots - this can occur close to polarity inversion lines on the source surface.}
\label{fig:step size}
\end{figure*}
To quantify how errors in field line tracing vary with tracing step size we produce error maps like Figure \ref{fig:field line map} for a range of spherical harmonic numbers. Figure \ref{fig:step size} shows how the maximum error in field line footpoint across a whole map varies with step size for different spherical harmonics. Errors with absolute values $>10^{\circ}$ are excluded from these plots, to exclude any points that are incorrectly traced near the polarity inversion line (see previous paragraph).

With decreasing magnetic field tracing step size, the maximum error drops off until reaching a plateau at a step size of $\sim$ 4. Because the field line integrator uses a 4th order Runge-Kutta method, it samples the field at 1/4 of the step size, so a levelling off of the error at a step size of around 4 is expected. Below this the integrator samples at sub-grid resolution, and the error is dominated by the error in the magnetic field solution itself. This result justifies our default choice of $\Delta s = 1$ for the other tests. Making the integration step size any smaller than $\Delta s = 1$ increases computation and memory use, but does not increase the accuracy at which the field lines are traced.

\section{Conclusions}
We have derived a set of analytical closed solutions for the PFSS equations when the input magnetic field is a single spherical harmonic (Section \ref{sec:pfss}), along with analytical equations for magnetic field lines traced through these solutions (Section \ref{sec:streamlines}). These solutions have then been used to test and quantify the accuracy of the numerical \pfsspy solver (Section \ref{sec:comparison}).

This set of tests reveal both the accuracy and limitations of the \pfsspy solver. The total open magnetic flux is systematically overestimated within the \pfsspy solver, but when the number of radial grid cells exceeds 40 is accurate to within 6\% of the true value for individual spherical harmonics (Figure \ref{fig:open_flux}). When tracing field lines from the source surface to the solar surface, \pfsspy is always accurate to within 0.5$^{\circ}$ with a small enough tracer step size (Figure \ref{fig:step size}), but large errors in field line tracing can occur near polarity inversion lines and at the poles (Figure \ref{fig:field line map}). Note that our tests use spherical harmonic solutions which are relatively smooth functions, so different tests are needed to understand how accurate the field line tracing is in areas of rapidly changing magnetic field, e.g., surrounding active regions.

These test problems form a solid benchmark for PFSS solvers and their ability to reproduce stellar potential magnetic fields accurately. We recommend that all PFSS solvers be tested against the analytical test problems presented here.
\begin{acknowledgments}
D.S.~and D.V.~are supported by STFC Consolidated Grant ST/S000240/1. D.V.~is supported by STFC Ernest Rutherford Fellowship ST/P003826/1.
\end{acknowledgments}

\bibliography{/Users/dstansby/Dropbox/zotero_library}
\bibliographystyle{aasjournal}

\end{document}